\documentstyle[prl, twocolumn, aps, psfig]{revtex}

\begin{document}
\draft
\preprint{}
\title{Revised Relativistic Hydrodynamical Model for Neutron-Star Binaries}

\author{G. J. Mathews}

\address{
University of Notre Dame,
Department of Physics,
Notre Dame, Indiana 46556}

\author{J. R. Wilson}

\address{
University of California,
Lawrence Livermore National Laboratory,
Livermore, California  94550}

\date{\today}
\maketitle
\begin{abstract}
We report on numerical 
results from a revised hydrodynamic 
simulation of binary neutron-star
orbits near merger.  We find that
the correction recently identified by Flanagan 
significantly reduces but does not eliminate the neutron-star
compression effect.
Although results of the revised simulations
show that the compression is reduced 
for a given total orbital angular momentum,
the inner most stable circular orbit
moves to closer separation distances.
At these closer orbits
significant compression and even collapse is still
possible prior to merger for a sufficiently soft EOS.
The reduced compression in the corrected simulation is 
consistent with other recent studies  of
rigid irrotational binaries in quasiequilibrium in which
the compression effect is observed to be small.
 Another  significant effect of this correction is
that the derived binary orbital frequencies are now in
closer agreement with post-Newtonian expectations.
\end{abstract}

\narrowtext

\section{INTRODUCTION}
In a recent paper  E.~Flanagan \cite{flanaganb} has pointed out
an inconsistency in the solution of the shift vector in
our previously reported 
numerical hydrodynamic simulations \cite{wm95,wmm96,mw97,mmw98}
 of binary neutron stars in quasiequilibrium orbits.  
In that paper it was suggested that this
may be the source of the controversial relativistic hydrodynamic
result that as the stars approach each other their interior density 
is observed to increase during numerical simulations.
 
In view of the controversial nature of the compression effect
\cite{contro}
and the fact that no other relativistic hydrodynamic 
treatments are yet capable of attacking this problem,
it is of course essential to incorporate this correction into
the previous hydrodynamic simulations to see if indeed this is the source
of the observed compression effect.  

In this short note we
report of the first results of simulations in which
this correction has been applied.  We find that
the correction does not eliminate the compression effect.  
It does, however, significantly 
diminish its magnitude at a given
angular momentum.  Perhaps more importantly, 
the correction  causes the orbital frequency of the
stars to increase such that the orbits
are in closer agreement with post-Newtonian expectations.  
At the same time, however, this causes the inner most stable
circular orbit (ISCO) to
move to closer separation distances and higher associated
fluid velocities.  The compression continues to scale with
the magnitude $U$ of the spatial component four velocity as
noted in \cite{mmw98}.  Hence, by the time of
inspiral and merger a significant compression effect is still 
possible.  For a sufficiently soft equation of state,
neutron stars in the mass range of observed neutron-star binaries
 still  collapse to individual black holes.

\section{The Correction}
The correction pointed out by Flanagan concerns the
the solution for the space-time components of the
ADM (3+1) metric
\begin{equation}
ds^2 = -(\alpha^2 - \beta_i\beta^i) dt^2 + 2 \beta_i dx^i dt + \gamma_{ij}dx^i dx^j~~,
\end{equation}
where the space-time components of the metric,
 $\beta^i$ are referred to as the shift vector.

We solve for the components of the shift vector by 
applying the ADM momentum constraint \cite{york79},
\begin{equation}
D_i(K^{ij} - \gamma^{ij}K) = 8 \pi S^j~~.
\label{evans37}
\end{equation}
Where $D_j$ is the three-space covariant derivative \cite{york79},
 and $S^i$ is the ADM three-momentum density.  

In our approximation scheme, we impose maximal-slicing 
[$Tr(K^{i j}) = 0$] and demand that the spatial
three metric $\gamma_{i j}$ be conformally-flat.
Under these constraints, the second term on the left hand
side of Eq. (\ref{evans37}) vanishes and we have,
\begin{equation}
D_i K^{ij} = 8 \pi S^j ~~.
\label{evans38}
\end{equation}
Ultimately, we can reduce equation  (\ref{evans38}) 
to a Poisson-like equation for the shift vector,
\begin{equation}
\nabla^2 \beta^i = 4 \pi \rho_\beta^i
-{\partial \over \partial x^i} \biggl({1 \over 3}
\nabla \cdot \beta\biggr) ~~.
\label{shift}
\end{equation}
The correction pointed out in \cite{flanaganb} is in the derivation
of the source density $\rho_\beta$. In \cite{wmm96} the 
contravariant spatial components of the four-momentum density
were utilized in place of the contravariant three-momentum
density.
[This is valid for the covariant components but not for the
contravariant components.]  Hence, in \cite{wmm96}
the source density was incorrectly written as,
\begin{eqnarray}
\rho_\beta^i  =& & 4\alpha \phi^4 S_i - 4 \beta^iW^2\sigma   \\
 & ~~+ & {1 \over 4\pi \xi}{\partial \xi  \over \partial x^j}
\biggl({\partial \beta^i \over \partial x^j}
+{\partial \beta^j \over \partial x^i}
-{2\over 3}\delta_{ij} {\partial \beta^k\over \partial x^k}\biggr)~~, \nonumber
\end{eqnarray}
where $\sigma = \rho + \rho \epsilon + P$ is the inertial mass density,
$\xi \equiv \alpha/\phi^6$, $ W \equiv \alpha U^t$, and 
\begin{equation}
S_i = \sigma W U_i~~.
\label{seq}
\end{equation}

The correct source 
should not have the second term on the right hand side.
This is a spurious term which came from an erroneous 
identification of the contravariant momentum density with the 
ADM contravariant three-momentum density.
The correct source density is,
\begin{equation}
\rho_\beta^i = 4\alpha \phi^4 S_i   
  +  {1 \over 4\pi \xi}{\partial \xi  \over \partial x^j}
\biggl({\partial \beta^i \over \partial x^j}+
{\partial \beta^j \over \partial x^i}
-{2\over 3}\delta_{ij} {\partial \beta^k\over \partial x^k}\biggr), 
\end{equation}
where $S_i$ is still given by Eq. (\ref{seq}).

This correction would have little effect on the final results
if the spurious $4 \beta^iW^2 \sigma$  term were small compared
to the term containing $S_i$.  That is not the case however. 
 In the simulations
of \cite{wmm96} it was observed that those two terms nearly
cancelled leading to small values for the $\beta^i$.
 With this cancellation no longer in effect, the source
for the shift vector equation 
is now a substantial quantity.  This has several implications
for the orbit dynamics which we now analyze.

\section{Results}

\subsection{$\Gamma = 2$ EOS}

As one way to identify the effect of this correction we have 
done calculations similar to the bench-mark calculation 
of \cite{mmw98}, i.e.~we have employed a simplistic $\Gamma=2$ polytropic equation of 
state (EOS),
$P = K\rho^\Gamma$, where $K= 1.8 \times 10^5$ erg cm$^3$ g$^{-2}$.
This gives a maximum  neutron-star mass
of 1.82 $M_\odot$.
The gravitational mass of a single $m_B = 1.625$ $M_\odot$ star 
in isolation is 1.51 $M_\odot$ and the central density is
$\rho_c =5.90\times 10^{14}$ g cm$^{-3}$.  
The compaction ratio for an isolated star with the
same numerical grid resolution (in terms of Schwarzschild
coordinates) is  $m/R = 0.14$.  [Note that  the central density
for these stars these stars
is slightly higher than 
quoted in \cite{mmw98} (for which 
$\rho_c = 5.84 \times 10^{14}$ g cm$^{-3}$). This is due to slight
changes in the finite differencing of the present calculation.]
The binary is taken to have
a fixed angular momentum $J = 2.5 \times 10^{11}$ cm$^2$
($J/M_G^2 = 1.27$ where M$_G = 2 m_G$).  In quasiequilibrium circular orbit
for this  angular momentum, the proper (coordinate)  separation distance 
between centers is 118 (102)  km with a frequency consistent with
post-Newtonian estimates.  The stars
relax to a nearly irrotational flow, and  the central density
only slightly increases  by 0.8\% to $5.95 \times 10^{14}$ g cm$^{-3}$
as summarized in Table  \ref{gam2}.

This very slight increase in central density is comparable
to the numerical accuracy of the hydrodynamic calculation which we estimate
to be $\sim \pm$0.5\%.
It is also consistent with the very small amount of compression noted 
in simulations of purely irrotational $\Gamma = 2$ stars
by several groups \cite{bonazzola99,mmwir,uryu} with a similar compaction ratio.
For example, in \cite{bonazzola99} stars with a similar compaction ratio
exhibit an increase in central density of 
0.1\% at this separation distance.  In \cite{mmwir} changes in central density 
 of order $ 0.1 \pm 0.5\%$ are consistent with the numerical results; and in
\cite{uryu}, stars with $m/R = 0.14$ show a maximum
central compression of about 2\% at a coordinate separation distance of $d/R_0 = 1.6$ 
in their notation.  A reasonable extrapolation of their figure 6 to our separation
 ($d/R_0 \approx 4.5$) suggests an increased central density consistent with our results.

\subsection{Realistic EOS}

In all of the above simulations  \cite{bonazzola99,mmwir,uryu},
it has been noted that the compression  effect dramatically increases
for an EOS with an increased compaction ratio and  closer orbits.
Hence, in the remaining discussion we consider the "realistic"
neutron star equation of state from \cite{wmm96,mw97} for which the compaction ratio
is much higher $m/R \approx0.25$.
For this set of calculations, we not only compare with previous
published results \cite{wmm96,mw97}, but also quantify the 
effect of the correction on the location of the ISCO  and
the orbital frequency for the binary.  

Consider  first the somewhat soft
 EOS with a critical
mass of $m_c = 1.575$ and stars with a baryon mass
of 1.548 $M_\odot$ corresponding to M$_G$ = 1.39 M$_\odot$
$\rho_c = 1.34 \times 10^{15}$ g cm$^{-3}$ in isolation.  
These parameters roughly correspond \cite{eosdiff} to
one of the simulations in Table 2 of Ref.~\cite{mw97}.
Table \ref{real2} compares differences between the previously
derived quantities \cite{mw97} and the corrected values for
the $J = 2.7 \times 10^{11}$ cm$^{2}$ run.  For the
uncorrected runs, this $J$ was the last orbit  
before the stars collapsed.   

Here, similar trends
to those noted in the $\Gamma = 2$ EOS are to be noted.
The correction significantly stabilizes the stars
for this $J$.  However, the 
amount of compression ($\sim 3\%$) for this EOS 
is greater than for the simulation in
Table \ref{gam2}.

In Table \ref{realeos} we show properties of corrected orbits
as the stars approach the ISCO.  For the uncorrected simulations
the ISCO for these stars with a stiff EOS (i.e no collapse)
 occurred at $J \approx 2.0 \times
10^{11}$ cm$^{2}$.  In the corrected runs 
the orbital frequencies are larger for the same fixed angular
momentum.  This moves the ISCO in to closer distances and smaller $J$.
Thus,  although the compression effect is less for a given $J$,
the stars reach higher velocities before the ISCO. This allows for
significant compression ($\sim$10\% in this case) before the 
ISCO.  Note, however,
that these stars do not collapse as they had done in the
uncorrected simulations.  

We did find, however, that collapse
could occur if the stars were increased in mass  from 
$m_G = 1.39$ to 1.44 M$_\odot$  ($m_B = 1.61$ M$_\odot$)
for this simulation. 
The results from this run are given in Table \ref{realeos2}.
Collapse of the stars  was observed to occur 
for very close orbits ($J = 1.85
\times 10^{11}$ cm$^{2}$) just before inspiral.
The coordinate separation between stars was only 2.4 times
the coordinate radii.  At  $J = 1.8
\times 10^{11}$ cm$^{2}$ the orbit still appeared stable though the
stars wanted to collapse.  In another calculation we used a softer EOS for which
$m_c = 1.54$ M$_\odot$ and $m_G = 1.40$, $m_b$ = 1.54  M$_\odot$.  For this case,
collapse occurred with $J = 2.0 \times 10^{11}$ cm$^{2}$.

 Thus, collapse may still be a possibility albeit for stars
close to the maximum mass of a soft EOS 
and for very close orbits.  Such is soft EOS is a reasonable
possibility.  For example, collapse would always occur
prior to inspiral for typical-mass neutron stars 
modeled with the EOS of Bethe and Brown \cite{bb95}.

\subsection{Angular Momentum at the ISCO}
Another significant effect
concerns the specific angular momentum
$J/M_G^2$ as the stars approach the final orbits.
In the previous results
\cite{eardley} the specific angular momentum at the ISCO
was significantly higher ($J/M_G^2 \approx  1.3$)
 than that of a maximally rotating Kerr black hole. 
This would imply complicated dynamics during inspiral before the 
stars could merge.
For the  corrected results, $J/M_G^2 = 1.03$ for the last computed  
stable orbit of the sequence in Table \ref{realeos}. 
For  $J/M_G^2 = 0.99$  the
orbit is unstable.  Thus, we expect that
when the stars begin to inspiral the specific angular momentum
is very near unity, $J/M_G^2 \approx  1.00$ 
and will become  $\le 1$ as the orbit plunges.  Hence,
the stars can immediately  spiral inward to form a Kerr
black hole near maximum rotation.  This has important
implications for the emergent gravity wave signal from the
subsequent ringing.  

%

\section{Discussion}
Perhaps the most significant effect concerns the orbital frequency.
The correction has caused the frequency to  increase to be much closer to
the expected post-Newtonian result (cf.~Table \ref{real2}).  It is easy to understand why
this is so.  In the hydrodynamic calculations, the orbital
frequency $\omega$ is determined by minimizing the average
coordinate three velocity, i.e.
\begin{equation}
\langle V \rangle \propto  \langle U - \beta - \omega \times r\rangle \approx 0~~.
\end{equation}
The vector quantity $\beta$ is of opposite sign to $U$
and is much larger in the corrected simulations.  This means
that a larger value for $\omega$ is required to
minimize the three velocity.  This remains true even though
the magnitude of $U$ for the binary slso slightly  decreases. 

A way of understanding the decrease in $U$ is to consider the balance between 
the relativistic analog of the centrifugal force and
the gravitational force  \cite{wmm96}.  The $\beta$ part of the
centrifugal force can be written as  $S_j {\partial \beta^j /\partial x^i}$. 
This centrifugal force consists of two parts.
One part scales as $U^2$ and one scales as $\beta U$.
Since the $\beta$ term is now much larger the magnitude of $U$
required to balance the gravitational force  becomes smaller.

Alternatively the reason for a decrease in $U$ while $\omega$ increases
can be seen from the definition of angular momentum.
In our simulations we constrain each orbit
to a specific value for the covariant z component
of the orbital angular momentum vector.
In the corrected simulations the shift vector 
contributes significantly to the covariant angular momentum,
\begin{equation}
J_z^{tot} \equiv J = J_z^S + J_z^\beta~~.
\end{equation}
where
\begin{equation}
J_z^S = \int \biggl(x S_y - y S_x \biggr) {\phi^2 \over \alpha} d^3 x~~,
\label{jzs}
\end{equation}
and 
\begin{equation}
J_z^\beta= \int \sigma_\beta \biggl[y \beta^x - x \beta^y \biggr] d^3 x~~,
\end{equation}
where
\begin{equation}
\sigma_\beta = \biggl(W^2 \sigma - \rho \epsilon - P + 
{1 \over W}\biggr) {\phi^6 \over \alpha}~~.
\end{equation}
In the corrected equations $J_z^\beta$ is a significant
fraction of the total angular momentum ($^<_\sim 1/2$, 
see Table \ref{real2}).  Thus, 
the corresponding momentum density contributions ($J_z^S$)
for a given total angular momentum
is less.  This constrains $U$ to be smaller in equation \ref{jzs}.

In \cite{wmm96,mw97,mmw98} it was noted that the increase in central
density scales with the four velocity for the binary as $\rho_c \propto U^4$.
It is note worthy  that the $U^4$ scaling with a similar proportionality 
constant still applies to the corrected results as shown in
Table \ref{realeos2}. The main difference, however, 
is that the values of $U$ are 
lower after the correction.
Hence, the compression is less for a given total $J$.

\section{Conclusion}

  In summary, we have applied the revision recently
pointed out by E.~Flanagan \cite{flanaganb} into a hydrodynamic 
simulation of binary neutron stars near the ISCO.
The effects of this correction have been analyzed.
Although the magnitude of the compression effect
is reduced for a given fixed total angular momentum, 
some compression 
effect remains.  The scaling of the compression with
the spatial components of the four 
velocity $U$ remains the same.  The main difference 
is that $U$ is smaller for a fixed
angular momentum.  This is at least in part
because a significant
fraction of  the covariant angular momentum now arises from 
the corrected shift vector term.  

 Another outcome is that the corrected  orbital 
angular frequencies
are higher than the uncorrected results.  This brings
the anticipated orbital and gravity wave frequences into
 closer alignment
with post-Newtonian estimates.  Another consequence
of the higher orbital angular frequencies is that
the ISCO for the binary moves in to closer separation
distances,  higher velocities, and higher orbital
angular momenta than previously estimated.  
This increases the compression
such that collapse to two black holes is still
possible before inspiral, although 
only for a softer EOS and/or higher mass.
The possibility that the EOS is this soft
is not yet ruled out 
\cite{bb95,lattimer,nsmass,nseos,schaab,salmonson99}.

Although we note here that some compression remains
in the revised hydrodynamics, its magnitude
is considerably reduced.  The question remains
then as to whether the remaining effect is real or
an artifact of the uncertainties \cite{mmw98,bonazzola99} introduced by
the conformally flat condition on the metric.  
The resolution of this question, however, will require
simulations in which the full Einstein dynamics
are included in the orbits.  We are currently developing
a perturbation expansion to examine this question.

 
Work at University of Notre Dame supported by NSF grant PHY-97-22086. Work at the 
Lawrence Livermore National Laboratory performed in part under the auspices of the 
U.~S.~Department of Energy under contract W-7405-ENG-48 and NSF grant PHY-9401636.

\newpage
\begin{table}
\caption{Comparison between corrected and uncorrected 
 results for
m$_B = 1.625$ M$_\odot$ ($m_G = 1.51$ M$_\odot$ in isolation)
 stars in a binary with
$J = 2.5 \times 10^{11}$ cm$^{2}$ and a  $\Gamma=2$ EOS.}
 \begin{tabular}{cccc}
 Quantity  & Isolated & Corrected & Uncorrected  \\
         & Star & Binary    & Binary \\
&&\cr
 \tableline
&&\\
$d_p$ (km) & $\infty$   & 118 & 138 \\
$\rho_c$ ($10^{14}$ g cm$^{-3})$  & $5.90$ & $5.95$ & $6.68$ \\
$(\delta \rho / \rho_0)$ & 0.  & $0.008$ & $0.14$ \\
\label{gam2}
\end{tabular}
\end{table}

\begin{table}
\caption{Comparison between corrected and uncorrected results
(from [4])  for
$m_B = 1.548$ M$_\odot$ ($m_G = 1.39$ M$_\odot$ in isolation)
stars in a binary with
$J = 2.7 \times 10^{11}$ cm$^{2}$ and a realistic  EOS.}
 \begin{tabular}{ccc}
 Quantity  &
Corrected & Uncorrected  \\
&&\\
 \tableline
&&\\
$d_p$ (km)   & $61$ & 68 \\
$\rho_c$ ($10^{15}$ g cm$^{-3})$  & $1.38$ & $1.98$ \\
$(\delta \rho / \rho_0)$   & $0.03 $ & $0.51$ \\
$\omega$ (rad sec$^{-1})$ &  1330 & 732  \\
$(\omega / \omega_{PN})$   & 1.01  & 0.66  \\
$U^2 = W^2 - 1$ & 0.0091  & 0.034  \\
$J_z^{S}$ ($10^{11}$ cm$^{2}$) & 1.4   & 2.7   \\
$J_z^{\beta}$ ($10^{11}$) cm$^{2}$ & 1.3 & - \\
$\alpha_{min}$  & 0.55  & .44  \\
\label{real2}
\end{tabular}
\end{table}

\begin{table}
\caption{Summary of corrected results for
m$_B = 1.548$ M$_\odot$ ($m_G = 1.39$ M$_\odot$ in isolation)
binary stars with a realistic EOS.}
 \begin{tabular}{ccccc}
 $J$ ($10^{11}$ cm$^{2}$) & $\omega$ (rad sec$^{-1}$) &  $d_p$ (km)  & 
$\rho_c$ (10$^{15}$ g cm$^{-3}$) &  $U^2$\\
&&\\
 \tableline
&&\\
$\infty$  & $0.0$ & $\infty$ &  1.34 & 0.0    \\
3.0  &  1025  & 72.0 & 1.37 &  .0080 \\
2.8  &  1190  & 65.6 & 1.375 &   .0085\\
2.6  &  1460  & 56.2 & 1.38 &   .0097\\
2.4  &  1660  & 51.4 & 1.40 &   .0115\\
2.2  &  2110  & 43.8 & 1.43 &  .0145\\
2.1  &  2425  & 42.0 & 1.44 &    .0165\\
2.0  &  2530  & 41.2 & 1.45 &    .0185\\
1.9  &  2750  & 39.0 & 1.455 &    .020 \\
1.8  &  3000  & 35.0 & 1.46 &    .023 \\
1.7  &  3450  & 34.0 & 1.47 &    .0245\\
1.65 &  4200  & 29.2 & 1.50 &    .0335\\
1.6   & (Inspiral) &      &     &   \\
\label{realeos}
\end{tabular}
\end{table}

\begin{table}
\caption{Summary of corrected results for
m$_B = 1.61$ M$_\odot$ ($m_G = 1.44$ M$_\odot$ in isolation)
 binary stars with a realistic EOS.}
 \begin{tabular}{ccccc}
 $J$ ($10^{11}$ cm$^{2}$) & $\omega$ (rad sec$^{-1}$) &  $d_p$ (km)  &
$\rho_c$ (10$^{15}$ g cm$^{-3}$) &  $U^2$\\
&&\\
 \tableline
&&\\
$\infty$  & $0.0$ & $\infty$ &  1.38 & 0.0    \\
3.0  &  1710  & 56.4 & 1.51 &  .013  \\
2.6  &  1800  & 52.0 & 1.55 &   .015 \\
2.4  &  2100  & 47.2 & 1.57 &   .018 \\
2.2  &  2400  & 44.2 & 1.60 &  .020 \\
2.0  &  3000  & 37.2 & 1.64 &    .030 \\
1.9  &  3600  & 29.8 & 1.72  &    .040 \\
1.85 (Collapsing) &  4500  & 23.6 & 4.05 &    .070 \\
\label{realeos2}
\end{tabular}
\end{table}


\begin{references}
\bibitem{eosdiff} The equation of state utilized in the
present simulation gives a gravitational mass which is differs by $< 1$\%
from the calculations of \cite{mw97}.  This difference is insignificant for 
purposes the present  comparison.
\end{references}
\end{document}